# Resource Usage Estimation of Data Stream Processing Workloads in Datacenter Clouds

Alireza Khoshkbarforoushha
Australian National University (ANU)
and CSIRO, Australia
a.khoshkbarforousha@anu.edu.au

Rajiv Ranjan
CSIRO, Australia
Australian National University
(ANU) raj.ranjan@csiro.au

Raj Gaire
CSIRO, Australia
raj.gaire@csiro.au

Prem P. Jayaraman
CSIRO, Australia
Prem.Jayaraman@csiro.au

John Hosking
Department of Computer Science,
University of Auckland
j.hosking@auckland.ac.nz

Ehsan Abbasnejad
Australian National University (ANU)
and NICTA
eabbasnejad@anu.edu.au

## ABSTRACT
Real-time computation of data streams over affordable virtualized infrastructure resources is an important form of data in motion processing architecture. However, processing such data streams while ensuring strict guarantees on quality of services is problematic due to: (i) uncertain stream arrival pattern; (ii) need of processing different types of continuous queries; and (iii) variable resource consumption behavior of continuous queries.

Recent work has explored the use of statistical techniques for resource estimation of SQL queries and OLTP workloads. All these techniques approximate resource usage for each query as a single point value. However, in data stream processing workloads in which data flows through the graph of operators endlessly and poses performance and resource demand fluctuations, the single point resource estimation is inadequate. Because it is neither expressive enough nor does it capture the multi-modal nature of the target data.

To this end, we present a novel technique which uses mixture density networks, a combined structure of neural networks and mixture models, to estimate the whole spectrum of resource usage as probability density functions. The proposed approach is a flexible and convenient means of modeling unknown distribution models. We have validated the models using both the linear road benchmark and the TPC-H, observing high accuracy under a number of error metrics: mean-square error, continuous ranked probability score, and negative log predictive density.

## 1. INTRODUCTION
Resource estimation underlies various workload management strategies including dynamic provisioning [28][34], workload scheduling [15], and admission control [4][35]. All these approaches possess a prediction module in common which provides estimations to determine respectively whether or not to add more resources, rearrange the order of query execution, and admit or reject a new incoming query[35].

Data stream processing workload is predominantly formed based on the job (i.e. continuous query) features registered and ordered in a stream processing framework and data arrival rate distribution models. When it comes to a cloud environment, the key to proper exploitation of cloud computing elasticity as a natural fit for handling uncertain data volume and velocity is having reliable information provided by query resource estimators where incoming queries schedule in different instance sizes based on resource demands. Therefore, having resource usage estimation for continuous queries is vital, yet challenging due to variability of the data arrival rates and their distribution models (e.g. Logistic, Rayleigh), variable resource consumption behavior of continuous queries, the need of processing different types of continuous queries, and uncertainties of the underlying cloud environment.

These complexities challenge the task of efficiently processing such streaming workloads on cloud infrastructures where users are charged for every CPU cycle used and every data bytes transferred in and out of the cloud. In this context, cloud service providers have to intelligently balance between various variables including compliance with Service Level Agreements (SLA) and efficient usage of infrastructure (and their cost) at scales where accounting for simultaneous peak workloads from many clients must be guaranteed without over-provisioning.

Recent work has studied SQL query resource estimation and run-time performance prediction using machine learning (ML) techniques [1][14][24], which treat the database system as a black box and try to predict based on provided training dataset. These techniques offer the promise of superior estimation accuracy, since they are able to account for factors such as hardware characteristics of the systems or interaction between various components. All these techniques approximate resource usage for each query as a single point value.

Unlike standard SQL queries that may (not) execute multiple times (each execution separate from previous one), continuous queries are typically registered in stream processing systems for a reasonable amount of time and streams of data flow through the graph of operators over this period. Rapidly time-varying data arrival rates and different query constructs (e.g. time and tuple-based windows) cause the resource demand of a given query fluctuate over time. To illustrate how streaming workload resource demands fluctuate by time, we picked and ran the following simple CurActiveCars [29] query from the linear road benchmark [2]:

> SELECT DISTINCT car_id
> FROM CarSegStr [RANGE 30 SECONDS];

Figure 1 (a) illustrates the CPU usage for this query against two different arrival rates: 500 msg/sec and 10,000 msg/sec. As we expect, the data arrival rates affect the stream processing system resource demand drastically over time. For example, the fitted Probability Density Function (pdf) to the CPU usage of a CurActiveCars query (Fig. 1 (b)), shows that even though the query is highly likely to consume between 20% to 35% CPU usage, we need to allow for possible peak demands (90% usage) to avoid a performance hit and its following SLA violations.

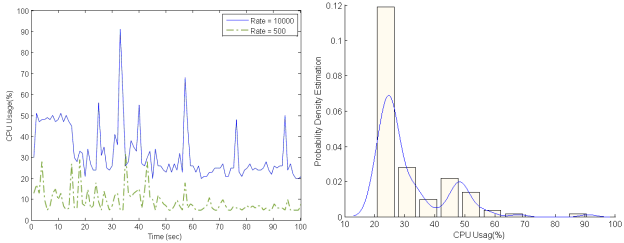

**Figure 1: (a) CPU usage of the query against 500 and 10k message per second arrival rates. (b) Normalized histogram and kernel density estimation (kde) fitted to CPU usage of CurActiveCars query against 10k data arrival rate.**

Under these circumstances, how can we address difficult questions such as "*How much Memory will the query require if the arrival rates double?*","*What would be the shape of CPU usage for more complex queries?*".

For problems involving the prediction of continuous variables, the single point estimation which is, in fact, a conditional average provides only a very limited description of the properties of the target variables. This is particularly true for data stream processing workload in which the mapping to be learned in multi-valued and the average of several correct target values is not necessarily itself a correct value. Therefore, single point resource usage estimation [1][14][24] will not be adequate for streaming workload, since it is neither expressive enough nor does it capture the multi-modal nature of the target data.

To overcome this issue, continuous queries and streaming workload resource management strategies require techniques that provide a holistic picture of resource usage as a probability distribution rather than simply a single most likely value. To achieve this, we propose a novel approach for resource usage estimation of data stream processing workloads. Our approach is based on the Mixture Density Network (MDN) [5], which approximates the probability distribution over target values.

Our approach combines the knowledge of continuous query processing with MDN statistical models. To do so, we first execute training query workload and record their resource usage values along with predefined query features. Secondly, we input the query features and data arrival rates to the MDN model. Finally, the MDN model statistically analyzes the feeding features' numbers and actual observation of the resource consumption of training data and predicts the probability distribution parameters (i.e. mean, variance, and mixing coefficients) over target values. Once the model is built and materialized, it can then be used to estimate the resource usage value of new incoming queries based on the query feature value set obtained from different sources (e.g. query statement, query plan) without executing the query.

To illustrate the gains possible by using the proposed approach, consider Figure 2 which displays a sample predicted pdf and actual CPU usage in terms of normalized histogram and fitted kernel density estimation (kde) for one of the experiments on linear road benchmark [2] queries in this paper. As we can see, the estimated pdf approximates the actual resource usage pdf closely. The predicted pdf provides a complete description of the statistical properties of the CPU usage through which we are not only able to capture the observation point, but the whole spectrum of the resource usage. In contrast, a best approximation from existing resource estimation techniques [1][14][24] merely provide the point which is visualized by solid vertical line through which, unlike the pdf, we are not able to directly calculate some valuable statistical measures such as variance, expectation, confidence interval about the target data.

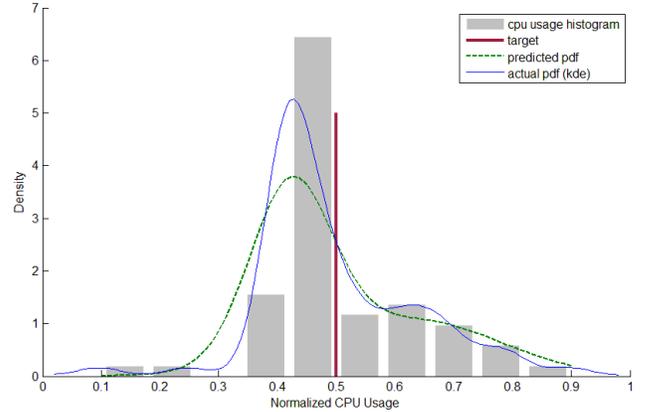

**Figure 2: sample conditional probability distribution predictions of CPU usage in case of registering NegAccTollStr [29] query. Actual pdf curve is a fitted kde function against the actual CPU usage that is used for clarity and comparison with prediction.**

To the best of our knowledge, this work is the first attempt toward resource usage estimation of data stream processing workloads over virtualized infrastructure. In summary, we make the following contributions:

- **Resource Modeling:** we have developed black-box models for predicting CPU and memory usage of centralized data stream processing workload based on continuous query features and data arrival rates. This is why the data stream processing workload behavior is predominantly the function of query plan complexity along with data arrival rate distribution models. It is important to note that the approach makes no assumption of the final shape of distribution (e.g. Weibull or Gamma distribution) which is the key in resource modeling of streaming workload as distribution models practically can be any shape and are application specific.
- **Evaluation:** we evaluate our models on a real stream processing system, both using the well-known linear road benchmark (LRB) [2] and TPC-H [30]. Our contribution here is first implementation and adaptation of the benchmarks, in particular, the TPC-H in a stream processing system. Second, building two workloads consist of more than 17200 and 8700 performance traces based on the LRB template queries[29] and the TPC-H respectively and evaluating the models. Both benchmarks contain queries with simple to highly complex execution plans that give us the confidence about the model robustness in the presence of wide range of continuous queries and system resource usage pattern.

In the reminder of this paper, we first discuss prior efforts in resource usage estimation of both online transaction processing (OLTP) and data stream processing workloads in Section 2. We next describe details of our approach in Section 3. We evaluate the performance of the solution using numerous experiments and report the results under different error metrics in Section 4. We finally present our conclusions from the work in Section 5.

## 2. RELATED WORK

There are roughly two line of related work; one directly investigates query performance prediction and another uses estimations for workload management. In this section, we will explore the second thread as well because it is highly relevant to our work and its exploration highlights some of the broken links between them.

Query processing run-time and resource usage estimation has been investigated in recent years. This line of work explores the estimation of run-time and also resource consumption of SQL queries in the context of both interleaved [1][14][24] and parallel execution [13][33] via black-box and white-box approaches. In the first approach, the underlying technique is based on statistical machine learning where the classifiers are built upon query plan features [14], operator level features, or both [1]. The role of database query plan optimizers in this approach is limited to provisioning of an early estimation of a set of features (e.g. operator input/output size, total plan cost) as input for training the model. The second approach [32] focuses on offline calibration of cost units of an optimizer cost model via profiling a set of calibration queries and online refinement of cardinality estimation using by sampling of the final query plan. The results show that using either a sophisticated ML algorithm or properly calibrating the optimizer lead to accurate query performance prediction.

ML algorithms compared to a white-box approach ease the task of cost model generation for increasingly complex data management systems as a wise selection and usage of various ML techniques captures implicitly the internal behavior of components and their interaction with operating system modules in terms of their resource footprint. This complexity is further intensified in cloud environments due to the heterogeneity, and diversity of resource types and uncertainties of the underlying cloud environment. However, building an accurate predictive model depends on the availability of training data which is a representative of the actual workload. In contrast, white-box approaches do not need training data and unlike sophisticated statistical models, they are easy to understand and intuitively fit into existing query optimization paradigms. Though they offer a higher extrapolation power, developing, enhancing and maintaining analytical models for ever-changing systems is quite challenging. These tradeoffs between two classes of models made researchers [25] build both white- and black-box models to predict the resource utilization (CPU, RAM, disk I/O, etc.) of the system for different transaction types in OLTP workload.

Resource sizing for data and stream processing systems typically follows one of two different coarse-grained and fine-grained approaches: i) analyzing the system performance and decision making based on the overall overload or under-load behavior of the system [21][8], and ii) query feature analysis and estimation [4][9][35]. The first approach formulates the auto-scaling requirement of the system as local [7], or global [19] thresholds, or build policies using a prediction model trained on historical feedback [21] in order to react to resource demands. In fact, the approaches are independent of the specifics of the query running in a data stream processing system or DBMS. However, the second strategy enjoys query performance estimations as a core indicator of workload behavior. In this category, the authors in [10] assume that there already exists a run-time estimation of queries so that they focus on how to use sharding or virtual partitioning to distribute load. Techniques used in [35][36] also share the same constraint. They simply assume that the pdf for the execution time of a query is already available for the service provider. Another related work in this category is reported in [4] where the authors proposed an input and query aware partitioning technique for load and QoS (e.g. latency) management. One of the key limitations of this approach is its reliance on meta-data information, expected to be provided by users, and the input rate estimation using time series forecasting. Predicting workload using a time series analysis based on the history or workload pattern is not feasible, since event rates and the data distribution for a data stream processing system usually change in an unpredictable way.

Based on the above discussions, readers may have noticed the broken link between the two threads of work. All the proposed techniques for resource estimation contemplate resource usage as a single point value prediction, whereas the techniques proposed in recent studies for data processing workload and SLA management [35][36] rely on the whole spectrum of resource usage. Because even in OLTP workload, queries with the same query time may follow different query time distributions [35]. This paper approaches this issue by proposing models which are able to provide a full pdf over the target domain, conditioned on the input.

## 3. THE PROPOSED APPROACH

The technique we describe in this paper combines the knowledge of continuous query processing with ML statistical models. In our approach, the continuous query feature set and data arrival rate distribution models form the input vector. This exploits an important observation, that data stream processing workload behavior is predominantly the function of query features and complexity along with data arrival rates.

The classic statistical ML techniques such as multilayer perceptron (MLP) are able to model the statistical properties of data generator. However, if the data has a complex structure, for example it is a one-to-many mapping, then these techniques are inadequate [26]. The scatter plot of CPU usage against average arrival rates in CurActiveCars query (Figure 3) illustrates the multi-valued mapping point in which for the same arrival rate such as 10k (msg/sec) there are many CPU usage values which range from 20 to 90 percent. This means that the conditional distribution (which can be visualized by considering the density of points along a vertical slice through the data) for many input value such as 10k or 9998 is multi-modal which can poorly represented by the conditional average. Therefore, we need a technique that is able to capture the multi-modal nature of the target data density function. Note that such a behavior in data stream processing workloads is unsurprising because the window construct has the potential to impose such significant variations in resource demands even we disregard arrival rates fluctuations or performance violation from other workloads.

Our approach employs MDN [5], a special type of Artificial Neural Network (ANN), in which the target (e.g. CPU usage) is represented as a conditional probability density function. The conditional distribution represents complete description of generator of data. An MDN fuses a mixture model with an ANN. We utilize a Gaussian Mixture Model (GMM) based MDN where the conditional density functions are represented by a weighted mixture of Gaussians. The GMM is very powerful means of modeling densities, since it is able to fully describe model by 3 parameters that determine Gaussians and their membership weights. From this density, we can calculate the mean which is the conditional average of the target data. Moreover, full densities are also used to accurately estimate expectation and variance that are two main statistics characterizing the distribution.

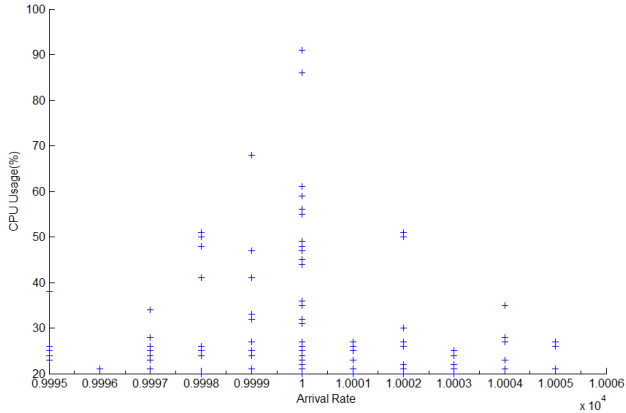

**Figure 3: CPU usage of CurActiveCars query against average arrival rates show the multi-valued mapping situation from the same input.**

Figure 4 gives an overview of our proposed approach in which collected query features from the continuous query language (cql) statement and query plan form the main input features of the model. In this process, the neural network is responsible for mapping the input vector $x$ to the parameters of the mixture model ($\alpha_i$, $\mu_i$, $\sigma_i^2$), which in return provides the conditional distribution. In fact, figure 4 shows a toy-example MDN with 2 components that takes a feature set $x$ of dimensionality 4 as input vector and provides the conditional density $p(t|x)$ over target $t$ of dimensionality 1.

A number of other approaches such as Conditional Density Estimation Network [27] and Random Vector Functional Link [22] are also available to estimate the conditional density function. The benefit of using MDN is due to its ability to model unknown distributions as exhibited by continuous queries and streaming workload. In addition, it has already been successfully applied in the other domains such as statistical parametric speech synthesis, finance, meteorology.

### 3.1 Resource Usage Modeling

Employing a machine learning approach requires fulfilling the following tasks: i) initial feature set identification, ii) model selection, iii) feature selection, and iv) training and testing. A key to the accuracy of a prediction model is the features used to train the model. We identify a set of potential features that would affect the performance (Task 1) and the query resource usage. The potential features or attributes are gathered by analyzing those considered in related work [14][1] and those we observed in various performance test analyses.

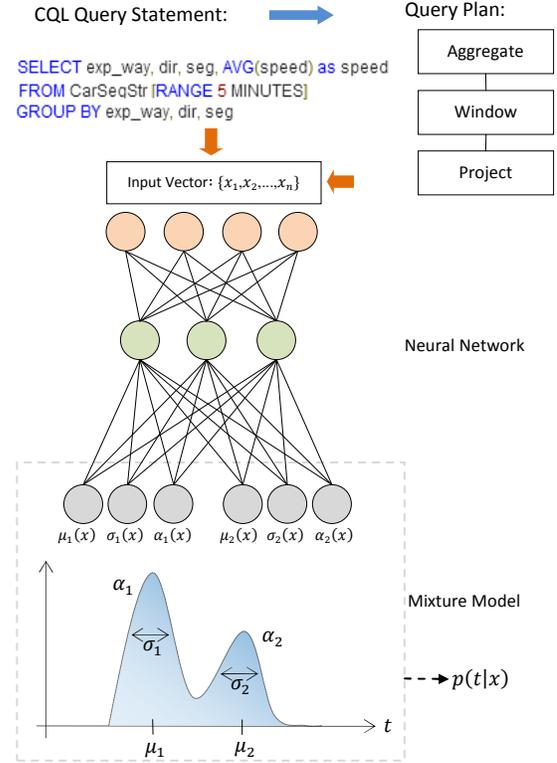

**Figure 4: Overview of the proposed approach using MDN for predicting the Gaussian Mixture Model parameters over target domain conditioned on input (i.e. query features).**

Intuitively, not all features have high-correlation with the target of the model and we need to select those with high predictive capability (Task 3). To this end, we use correlation-based feature subset selection method [20] along with best-first search for identifying the most effective attributes from feature vector space.

**Table 1: Feature input for training model.**

| Feature Name | Description | Collection Source |
|---|---|---|
| **avg_arrival_rate** | Average arrival rate of tuples (tuple/sec) to query | Load Generator Log/ Distribution Model |
| **stream_no** | Number of data stream sources | Query statement |
| **subquery_no** | Total number of nested sub-queries | Query statement |
| **agg_func_no** | Number of aggregation functions | Query statement |
| **join-predicate** | Number of join predicates | Query |

| | in query | statement |
|---|---|---|
| **project_size** | Projection size of query | Query statement |
| **equ_select_predicate** | Number of equality selection predicate | Query statement |
| **inequ_select_pridicate** | Number of non-equality selection predicate | Query statement |
| **agg_column_no** | Number of columns involved in GROUP BY clause | Query statement |
| **opt_type_count** | The number of each operator type in query plan | Query plan |
| **win_type_size** | The size of windows which is either time unit (sec) in time window or tuple unit (number) in tuple window type | Query Statement |
| **win_type_slide** | The sliding value of the window type | Query Statement |

Table 1 lists the feature set used as an input to the model. The attributes are extracted from multiple sources such as query statement text (e.g. window type and size), distribution model (e.g. arrival rate), or query plan (e.g. operator type).

Note that the above list is further customized based on the target of prediction, because they have different impact respecting to CPU and memory as resource modeling target. A feature that tremendously correlates memory consumption might have no correlation with CPU usage. For example, the feature selection task shows that the window size has an insignificant effect on CPU usage prediction, while it affects the memory usage prediction drastically.

### 3.2. Mixture Density Networks

The combined structure of feed-forward neural network and a mixture model form an MDN. In MDN, the distribution of the outputs $t$ is described by a parametric model whose parameters are determined by the output of a neural network, which takes $x$ as inputs. Specifically, an MDN maps a set of input features $x$ to the parameters of a GMM (i.e. mixture weights (or coefficients) $\alpha_i$, mean $\mu_i$, and variance $\sigma_i^2$) which in turn produces the full pdf of an output feature $t$, conditioned on the input vector $p(t|x)$. Thus, the conditional density function takes the form of GMM as follows:

$$p(t|x) = \sum_{i=1}^{M} \alpha_i(x)\varphi_i(t|x) \quad (1)$$

where M is the number of mixture components, $\varphi_i$ is the *i*th Gaussian component's contribution to the conditional density of the target vector $t$ as follows:

$$\varphi_i(t|x) = \frac{1}{(2\pi)^{c/2}\sigma_i(x)^c} exp\left\{-\frac{\|t - \mu_i(x)\|^2}{2\sigma_i(x)^2}\right\} \quad (2)$$

The MDN approximates the GMM parameters including mixture coefficients $\alpha_i$, mean $\mu_i$, and variance $\sigma_i^2$ as:

$$\alpha_i = \frac{\exp(z_i^\alpha)}{\sum_{j=1}^{M} \exp(z_j^\alpha)} \quad (3)$$

$$\sigma_i = \exp(z_i^\sigma) \quad (4)$$

$$\mu_i = z_i^\mu \quad (5)$$

where $z_i^\alpha$, $z_i^\sigma$, and $z_i^\mu$ are the outputs of the neural network corresponding to the mixture weights, variance, and mean for the *i*th Gaussian component in the GMM, given $x$ [5]. To constrain the mixture weights to be positive and sum to unity, *softmax* function is used in Eq. (3) which relates the output of corresponding units in the neural network to the mixing coefficients. Likewise, the variance parameters (Eq. 4) are related to the outputs of ANN which constrains the standard deviations to be positive.

The objective function for training the MDN is to minimize the negative log likelihood (NLL) of observed target data points given to mixture model parameters:

$$E = -\sum_n ln\left\{\sum_{i=1}^{M} \alpha_i(x^n)\varphi_i(t^n|x^n)\right\} \quad (6)$$

Since the ANN part of the MDN provides the mixture model parameters, the NLL must be minimized with respect to the network weights. In order to minimize the error function, the derivatives of the error function with respect to the network weights are calculated. Specifically, the derivatives of the error are calculated at each network output units including the priors, means and variances of the mixture model and then propagated back through the network to find the derivatives of the error with respect to the network weights. Therefore, standard non-linear optimization algorithms such as *scaled conjugate gradients* can be applied to MDN training.

Once an MDN is trained with regard to normal precautions of over-training and local minima, it can predict the conditional density function of CPU and memory, conditioned on the incoming query features and data arrival rate distribution model.

### 4. EXPERIMENT

To test if the streaming workload resource usage prediction can be successfully carried out in practice through an MDN technique, experiments needs to be designed, constructed, executed, and reported. This section explores the steps followed and the results obtained from the experiments conducted to evaluate the accuracy of our approach.

### 4.1 Experimental Setup

Performance analysis of stream processing system in the public cloud [8] already shows the variation of throughput and latency is trivial and the measures are relatively stable. We conduct our experiment in a private cloud to test the accuracy of estimations in presence of any possible performance variations.

Two virtual machine (VM) instances, one for load generation and another as a host for the stream processing system, were employed. The stream generator system was a m1.medium size instance with 4GB RAM, 2 VCPU running Ubuntu 12.04.02 Server 64b. All queries were executed on m1.large instance size with 8GB RAM, 4VCPU, and the same OS. The hypervisor is KVM, and the nodes are connected with 10GB Ethernet. In our cloud each physical machine has 16 cores of Intel(R) Xeon(R) CPU 2.20GHz with hyper threading enabled which the OS sees as 32 cores (CPU threads). Therefore, 4 VCPU map to 4 CPU threads and 2 full CPU cores.

### 4.1.1 Dataset and Workload

To validate our approach we implemented both the linear road benchmark [2] and slightly modified TPC-H in a commercial centralized stream processing system X.

**LRB Workload:** the Linear Road Benchmark has primarily been designed for comparing performance characteristics of streaming systems. It contains 20 queries with different level of complexities in terms of execution plan. We treated them as template queries. Excluding the ad-hoc query answering set reduced them to 17 template queries. Various arrival rates from 100 to 100k tuples per second along with random substitution of window size from 1 to 900 sec (1-100 row resp.) in the time window (tuple window resp.) resulted in 17289 execution traces (see Appendix A).

To generate data streams, 500MB data (i.e. 3 hours simulated traffic management data) was fed into the streaming system using the system's built-in load generator which played the role of data driver in the linear road benchmark. Each query was registered and logged for more than 3× of its window size to capture the impacts of time windows on resource consumption properly.

**TPC-H Workload:** In contrast to the LRB workload, TPC-H has primarily designed for DBMSs, though it has also been used in stream processing research[12][30]. In this context, each relation is considered as a data stream source and the tuples are sent toward stream processing engine over the network using a load generator. Therefore, each registered query references a subset of the relations in the input over time.

We created 0.1GB TPC-H database using *DBGen* tool as per the specification. To keep the overall experimentation duration under control we did not use larger database size (e.g. 1GB) because the tables are in fact the stream source material in our experiment and we have to send each tuple over the network. Quite simply, in 1GB database size, LINEITEM table has 6001215 tuples and even with the 5000 tuple/sec rate, it takes more than 20 minutes to send all the tuples over the network. With current hardware, this rate is maximum consumption rate for queries without any join such as Q1 and Q6. This rate drops to less than 200 for Q8 with 7 data stream sources.

As the system X does not support correlated sub-queries we forced to exclude templates 4, 11, 15-18, 20-22. We generated 35 executable query texts using *QGen* based on the remained 13 TPC-H templates queries.

Furthermore, we slightly modified these queries for our system to make them compatible with the stream processing context. One of the key changes was adding a time window for each stream sources to let queries show upper bound of CPU and memory usage. Moreover, some query semantics require that tuples not to leave the time window until a certain period of time to be able to produce meaningful results. In other words, we needed to keep the first tuple that enters the time window until the load generator reads and sends the last tuple from the relation source. To this end, we set the time window range to the value of $S$ if load generator needs $S$ seconds to read and send all the tuples.

The load generator was not allowed to send duplicate tuples. In addition, relations have different cardinalities so that in case of multiple stream sources in one query, we set all the time windows to the biggest one. This let the relation at time $t$ consists of tuples obtained from all elements of stream up to $t$. For example, in a join between LINEITEM (~600k tuples) and NATION (25 tuples) streams, the latter requires as big time window as the former to let elements remain in the window until the last tuple from LINEITEM stream enter the window for processing.

The 35 generated executable queries were registered separately in the system X and their performance measures against the fluctuating arrival rates were logged. The obtained workload consists of 8783 execution traces (see Appendix B).

Performance measures of interests (i.e. CPU and memory) were collected using the *dstat*[1]. This is a lightweight python-based tool that collects OS and system statistics passively, without affecting the performance. To guarantee healthy and repeatable data gathering, the execution traces of all queries were collected several times. Moreover, all queries were run sequentially with cold start making sure the buffers were flushed and we had a fresh JVM.

There is no documentation relating to the distribution model that the load generator follows, though these generators typically try to reach the specified rate, while adjusting rates based on engine consumption rate with the aid of a thread sleep function. This means, a query (especially complex ones) might be able to consume only 100 tuple per second even when we set the load generation rate to 200 tuple per second. Thus, a few seconds after commencement the buffer of the stream processing engine is full so that load generator thread sleeps for a few milliseconds to allow the consumer to exhaust the queue. This situation inherently emulates different load generation distribution, for example for rate 100 tuple per second the distribution model is t location-scale, whereas for rate 50k and 100k the distribution is more fit to Weibull and generalized extreme value distribution, as shown in Fig. 4.

### 4.1.2 Training and Testing Settings

To assess how the result of a predictive model would be generalized to an independent unforeseen data set, we divided the LRB workload randomly into training and testing dataset with 66% and 34% percentage split rates respectively. For TPC-H workload we used *k-fold cross-validation*. As regard to the workload size, *2-fold cross-validation* (also called holdout method) is used to learn and test parameters. For each fold, we randomly assigned data points to two equal size sets *ds1* and *ds2*. To do so, we shuffled the data array and then divided it in to two arrays. We then trained on *ds1* and test on *ds2*, followed by training on *ds2* and testing on *ds1*.

For conducting training and testing, we used a Netlab toolbox [26] which is designed to provide the central tools necessary for the simulation of theoretically well founded neural network algorithms and related models and in particular MDN. The implemented MDN model uses a MLP as a feed forward neural network, though in general any non-linear regressor can be utilized.

Before training and testing, all the training and testing workload was normalized. The input features were normalized using z-score, while the output features were normalized to be within 0.1-0.9 based on min-max normalization.

There are a number of hyper-parameters including the number of Gaussian components, number of neurons in MLP, etc. that need to be specified beforehand. We tried several settings and assessed the trade-off between accuracy, training time and overhead. Based on the obtained results, we conclude that a GMM with 3 components, 5 neurons per feature in the input vector, and normally more than 100 training cycles provides acceptable accuracy within a tolerable overhead.

---
[1] http://dag.wiee.rs/home-made/dstat/

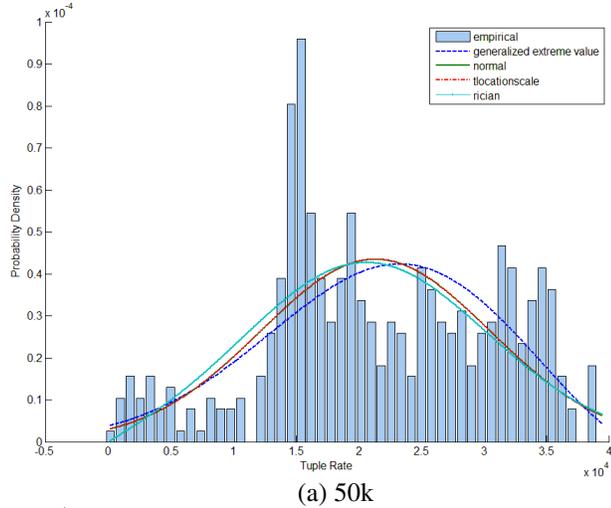

(a) 50k

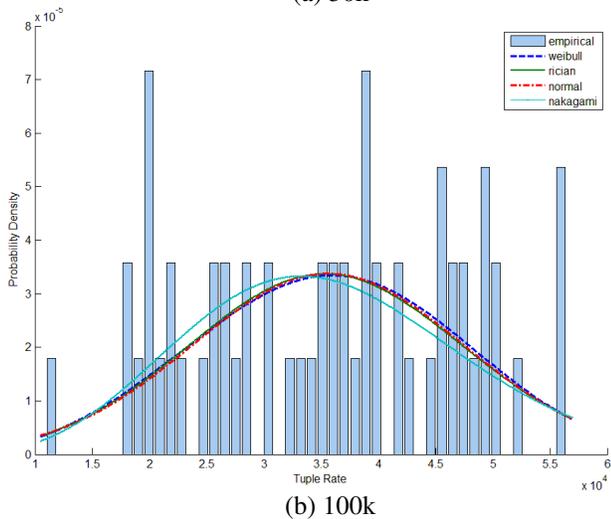

(b) 100k

**Figure 4: Best fit of sent tuple per second against different distribution models. The figures contain probability density of average tuple sent per second for the speed rate of 50k and 100k for two different queries.**

As discussed earlier, the loss function for MDN learning is NLL (Eq. 6). We have trained and tested the model over both the LRB and the TPC-H datasets for 100 iterations each containing more than 20 training cycles. Fig. 5 and 6 show the decaying NLL error function for the training and testing the LRB and the TPC-H datasets respectively. We had also used other settings (e.g. 4000 training cycles) in the test, however NLL losses were negligible.

As we can see in Fig. 5, the memory training and testing process has faster NLL loss. One reason for this behavior is that the current input features for training memory prediction model has a better predictive power compared to the CPU feature sets which lets the model to learn and generalize to independent dataset faster.

Another reason is that the input feature vector for memory has 14 attributes compared to 21 for the CPU. Combined with a smaller neural network size enables it to converge quicker in the same number of iterations. We had the same observation in the training and testing on the TPC-H benchmark, even though in these experiments both CPU and memory trained on the same set of input features. Unsurprisingly, the larger training set size of the LRB along with the less complexity of query execution plans make the model to be built and converged to lower NLL values compared to TPC-H workload.

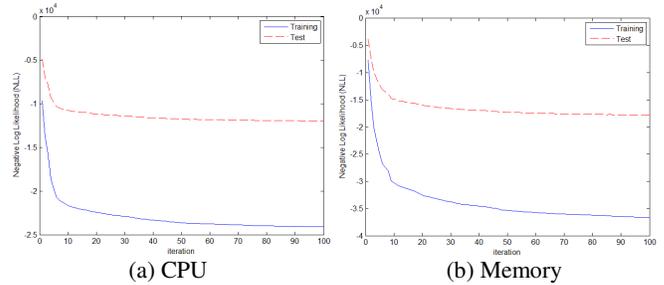

(a) CPU  (b) Memory

**Figure 5: Negative Log Likelihood (NLL) error of the training and testing on the LRB dataset in 100 iterations.**

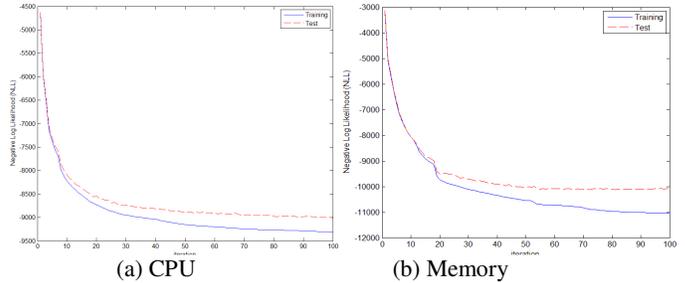

(a) CPU  (b) Memory

**Figure 6: NLL error of the training and testing on the TPC-H dataset in 100 iterations.**

### 4.2 Evaluation: CPU and Memory Usage

To determine if a probabilistic model performs well we must set the goals of using such a model [6] because if, for example, a trained MDN assigns some probability to the actual observation, we should be able judge whether the prediction is accurate or wrong. Therefore, in the following subsection we first set the goals and then define appropriate metrics for evaluation.

#### 4.2.1 Error Metrics

In essence, the goal of a probabilistic prediction is to maximize the sharpness of the predictive distributions subject to calibration [17]. Sharpness refers to the concentration of the predictive distributions. Calibration refers to the statistical consistency between the pdfs. Our objective is to predict calibrated pdfs that closely approximate the region in which the target lies with proper sharpness. To this end, the continuous ranked probability score (CRPS) [17] is a proper metric to evaluate the accuracy of pdfs. The CRPS takes the whole distribution into account when measuring the error.

$$CRPS(F,t) = \int_{-\infty}^{\infty} [F(x) - O(x,t)]^2 \, dx \quad (7)$$

where $F$ and $O$ are the cumulative distribution function of prediction and observation distributions, respectively. $O(x,t)$ is a step function that attains the value of 1 if $x \geq t$ and the value of 0 otherwise. Therefore:

$$CRPS(F,t) = \int_{-\infty}^{x_0} [F(x)]^2 \, dx + \int_{x_0}^{\infty} [F(x) - 1]^2 \quad (8)$$

To calculate CRPS both the prediction and the observation are converted to cumulative density functions. The CRPS compares the difference between cumulative distributions of prediction and observation as given by the hatched area in Fig 7. It can be seen that the area gets smaller if the prediction distribution concentrates probability mass near the observation, i.e. the better it approximates the step function. Moreover, the small CRPS value shows that the prediction captures the sharpness of prediction accurately.

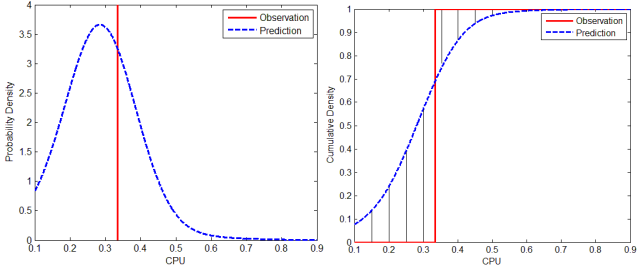

**Figure 7: (a) predicted pdf and the observation (b) schematic sketch of the CRPS as the difference between cdfs of prediction and observation.**

After calculating the CRPS for each prediction, we need to average the values to evaluate the complete input set:

$$CRPS = \frac{1}{n} \sum_{i=1}^{n} CRPS(F_i, t_i) \quad (9)$$

The CRPS generalizes the mean absolute error, thereby provides a direct way of comparing various deterministic and probabilistic predictions using a single metric formula [17].

We are also interested in evaluating the spread of predictive density in which our targets lie. The average negative log predictive density (NLPD) [18] error metric is used for testifying this aspect, although unlike CRPS it is not sensitive to distance:

$$NLPD = \frac{1}{n} \sum_{i=1}^{n} -\log\bigl(p(t_i|x_i)\bigr) \quad (10)$$

where $n$ is the number of observations. The NLPD evaluates the amount of probability that the model assigns to targets and penalizes both over and under-confident predictions. Both CRPS and NLPD are proper measures, meaning that they reward honest assessments [16].

The last metric we applied is the mean-square error (MSE) of predictions:

$$MSE = \frac{1}{n} \sum_{i=1}^{n} (t_i - m_i)^2 \quad (11)$$

where $m$ refers to the median of the pdfs as point predictions for the MDNs.

### 4.2.2 Evaluation Results

As mentioned earlier, there are a number of hyper-parameters including GMM components, number of neurons in MLP, etc. that need to be specified beforehand. To this end, we tested different MDN architecture including 3, 5, and 8 mixture components. Moreover, for each workload we ran numerous experiments at different settings as shown in Table 2 and 3. For the LRB workload we changed training cycles and number of neurons per features, while the train and test datasets kept unchanged throughout the experiments. Further, to see generalization power of the model, we applied 2-fold cross-validation on TPC-H workload, though the network settings were constant. The results for LRB and TPC-H workloads under CRPS, NLPD, and MSE metrics along with the selected architectures are shown in table 2 and 3 respectively. The variances over the results obtained in different settings are also shown underneath each error score.

**Table 2: CRPS, NLPD, and MSE error scores of trained MDN for the LRB workload.**

| Model | Resource | Arch. | CRPS | MSE | NLPD |
|---|---|---|---|---|---|
| MDN | CPU | 3 mix | **0.102** 0.0001 | 0.056 0.0005 | -0.264 0.405 |
| | | 5 mix | 0.128 0.0001 | 0.096 0.001 | -0.339 0.365 |
| | | 8 mix | 0.113 **0.00009** | **0.043** **0.0001** | **-0.865** **0.083** |
| | Memory | 3 mix | 0.065 0.0005 | 0.073 0.0008 | -0.496 0.463 |
| | | 5 mix | **0.053** **0.00004** | 0.066 **0.0002** | **-1.465** 0.09 |
| | | 8 mix | 0.065 0.0002 | **0.046** 0.002 | 0.075 **0.067** |

**Table 3: CRPS, NLPD, and MSE error scores for the TPC-H workload.**

| Model | Resource | Arch. | CRPS | MSE | NLPD |
|---|---|---|---|---|---|
| MDN | CPU | 3 mix | 0.171 0.006 | **0.009** **5E-07** | **-1.2** **0.02** |
| | | 5 mix | 0.16 **0.0018** | 0.02 0.0012 | -0.98 0.1 |
| | | 8 mix | **0.154** 0.002 | 0.02 0.005 | -0.9 0.057 |
| | Memory | 3 mix | **0.055** **0.00001** | 0.019 **8E-06** | **-1.09** **0.011** |
| | | 5 mix | 0.092 0.0004 | 0.094 0.001 | -0.91 0.038 |
| | | 8 mix | 0.097 0.0004 | **0.1** 0.001 | -0.67 0.02 |

All three metrics are negatively oriented scores, therefore the smaller value the better. In both workloads the error numbers are small enough that suggest the better accuracy of the model.

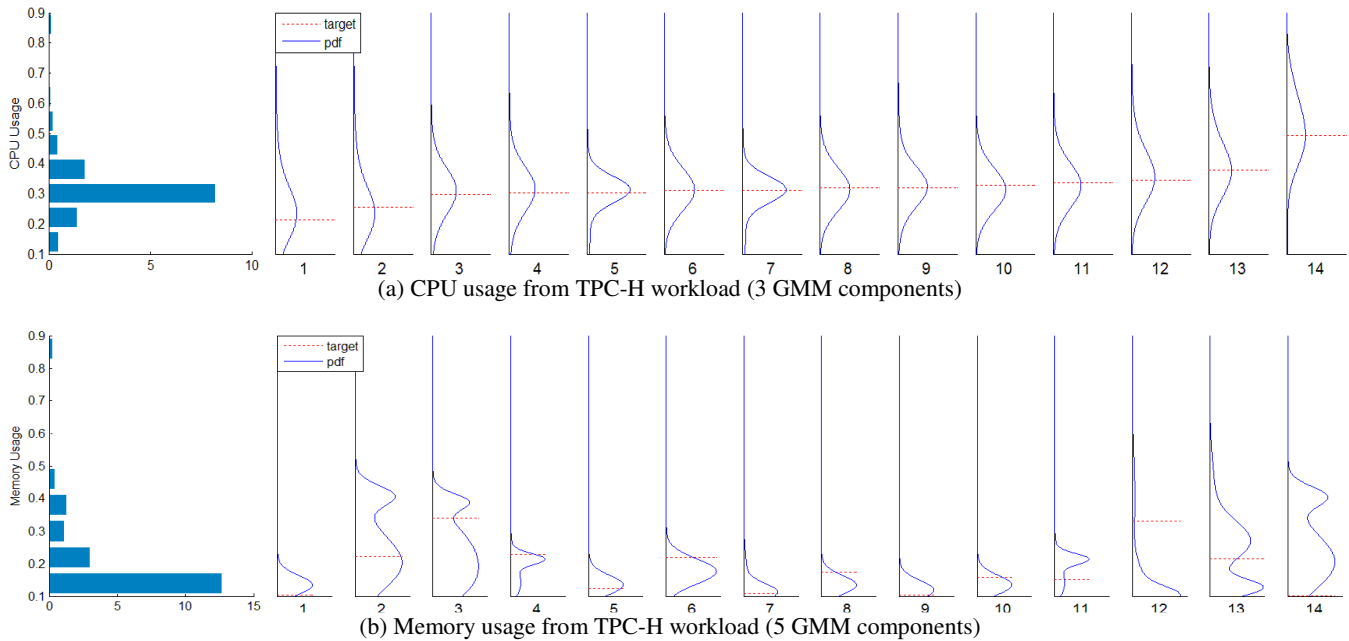

**Figure 8: sample pdf predictions of CPU and memory usage from TPC-H workload. The pdfs were selected randomly from test dataset which means they might belong to different queries and inputs.**

In LRB workload, sophisticated MDN architecture led to better results particularly under MSE and NLPD measures, though the improvement is not significant considering more components impose extra training times and overhead. In contrast, TPC-H workload shows slightly worse results as the architecture become more complex.

The overall results for both workloads are roughly similar; however the LRB workload shows less CRPS errors in both CPU and memory resources. There are two reasons for this: the complexity of queries and training data size. Specifically, TPC-H workload is by far complex than LRB as the query templates combine complicated query plans with various data sources. Although the LRB workload has wide complexity range of queries, all deal with one data stream. Moreover, the bigger workload size of LRB in training phase led to more accurate predictions.

### 4.3 Prediction Utilization

To provide a cloudless picture of how the provided prediction could be utilized and employed, we have visualized some sample conditional probability distribution predictions from test set of TPC-H workload. Figure 8 plots 14 random sample predicted pdfs for CPU and memory in which they were selected from MDN with 3 and 5 GMM components respectively. The histograms of the actual CPU and memory usage of the whole test dataset were also depicted to allow us to see the prediction success and failures explicitly. Each pdf may (not) belong to different queries as they were selected randomly from the test datasets. This means, pdfs conditioned on different inputs. The dotted vertical line in each digarm shows the observation value.

As both the figures 8(a) and 8(b) show, the pdfs successfully approximate the resource usage distribution which is respectively within the range [0.1, 0.6] and [0.1, 0.5] for CPU and memory usage. The models for CPU and memory resource usage above the values 0.6 and 0.5 are much more uncertain. In other words, the tendency of all CPU and memory pdfs are to the right hand side of diagram and this is consistent with the actual resource usage (i.e. plotted histograms) in which we hardly face resource demand above 0.5. The pdfs 2, 3, 13, and 14 unlike the others are bimodal in which two kernels have comparable priors. This means, the trained MDN is able to capture multi-modal nature of the target data as well.

These sample pdfs also confirm that the MDN is a useful classifier in the classic point estimate sense. As regard to this point, the CPU pdfs compared to memory figures perform better as the sharpness and the spread of predictive density cover the target properly. The memory pdfs particularly number 8, 11, 12, and 14 failed at accurate prediction of the target values, though except number 12 they nearly locate the spread of distributions.

This type of prediction gives the resource and workload manager a concise yet lucid means of workload behavior interpretation which is crucial for a number of applications including run-time performance isolation, SLA and billing management, diagnosis/performance inspection [25].

Upper and lower bounds of resource usage make the performance isolation task easy. For example, predictions in all figures capture the dominant CPU and memory usage precisely where the probability of available room for new queries allow an admission control (as a workload management strategy) to make a wise decision about whether to register a new query or not. SLA specifications and billing management also become more applicable and reliable for both clients and providers when there is an initial measure about the actual contribution of each workload to the overall resource consumption. Specifically, such prediction will be helpful in meeting SLA requirements that are expressed as high level QoS metrics (response time) but are directly dependent on low level resource QoS metrics (I/O, CPU, memory). When it comes to performance inspection, diagnosing abnormal behavior based on the predicted numbers is also viable. For example, Fig. 8 (b) reports that in these set of queries we will not face peak memory usage (>0.5) very often, hence superior peak memory usage in the live environment acknowledges the presence of a fault in the query processing process (e.g. VM performance issues, memory leak in stream processor) long before.

### 4.4 Constraints and Limitations

There exists no standard query language for continuous queries, though initial steps have been taken [23]. This issue leads to inconsistencies even in common capabilities among different SPEs where time-based windows, for example, are defined and specified differently and in some cases the underlying semantics of such common features are different [11] as well. However, there are a number of key operators (e.g. select, project, filter, time-based windows) that are common in all continuous query languages and most notably some differentiation in producing final results is negligible in terms of resource consumption patterns.

This non-standardization of continuous query language constrained us to select from the fixed collection of common operators that are supported by almost all continuous query languages so that the output model could be general enough to be applicable to all stream query languages. The provided operator list thus satisfies one more salient aspect, i.e., having sufficient constructs to be able to express the semantic foundations of continuous queries in a wide variety of stream applications.

In this paper we did not consider resource consumption estimation in the presence of concurrent executions and combined workload. This problem has become very important recently via introduction of new distributed real-time stream processing frameworks, but is left for subsequent work.

We also delimited our investigation to CPU and memory usage of centralized streaming system against registered queries. Though other resources such as disk I/O and Network along with the CPU and memory constitute the basic auto-scaling policies of public cloud providers in practice, continuous query computations are typically CPU or memory intensive jobs. Therefore avoiding performance hit is heavily dependent on the availability of these two resources. In other words, these resources are the important performance metrics for sizing and resizing of resources and scheduling of new queries in newly fired virtual machines.

### 5. CONCLUSION AND FUTURE WORK

In this paper we presented a novel approach for streaming workload resource usage estimation. Our approach shows acceptable prediction accuracy in the face of changes in distribution model of data arrival rate, complex queries with unknown resource usage probability density. To achieve these gains, our approach combined the knowledge of continuous query processing with mixture density networks that approximates conditional probability distribution density of resource usage, conditioned on query features and data arrival rate distribution model.

We evaluated the approach on a real stream processing system, both using the linear road benchmark and TPC-H. The obtained results under CRPS, NLPD, and MSE measures confirmed that our approach is able to quantify the uncertainty of streaming workload performance accurately. We believe that these models have a potential to become an integral component of the automated workload management systems for scheduling big data applications over private or public cloud resources.

A natural direction for follow-up work will be investigating how to deploy the proposed approach in a workload management strategy or how the prediction result as an input to workload management is reflected in SLA specifications. Moreover, to control the complexity of model and also identify the key features affecting streaming workload behavior, our approach covers continuous queries in the absence of concurrent executions and resource sharing. Therefore, future attempts apart from improving the accuracy of the current study with respect to the presence of resource sharing, could be targeted to parallel distributed stream processing workloads.

# APPENDIX
## A. LRB WORKLOAD DETAILS

This section presents details of the LRB workload generation. Different arrival rates and time/tuple window size for the queries in the linear road benchmark result in 17289 execution traces. It is important to note that, these two features were selected because their modification does not affect the query semantics, even though provides different selectivity ratio for query. Table 4 denotes the substitutions and number of execution traces collected for each query.

**Table 4: 17 template queries from linear road benchmark constitute workload traces for model training and testing.**

| Query No. | Query Name | Substitutions | No. of Traces |
|---|---|---|---|
| 1. | CarSegStr | Arrival Rate | 463 |
| 2. | CurActiveCars | Arrival Rate, Time Window Size | 8769 |
| 3. | CurCarSeg | Arrival Rate, Tuple Window Size | 404 |
| 4. | CarSegEntryStr | Arrival Rate | 130 |
| 5. | TollStr | Arrival Rate | 387 |
| 6. | SegAvgSpeed | Arrival Rate, Time Window Size | 3688 |
| 7. | SegVol | Arrival Rate | 157 |
| 8. | SegToll | Arrival Rate | 393 |
| 9. | AccCars | Arrival Rate, Tuple Window Size | 349 |

| 10. | AccSeg | Arrival Rate | 160 |
| --- | --- | --- | --- |
| 11. | AccNotifyStr | Arrival Rate | 158 |
| 12. | AccAffectedSeg | Arrival Rate, Time Window Size | 898 |
| 13. | CarExitStr | Arrival Rate | 149 |
| 14. | NegTollStr | Arrival Rate | 383 |
| 15. | NegAccTollStr | Arrival Rate | 189 |
| 16. | AccTransStr | Arrival Rate | 308 |
| 17. | AccBal | Arrival Rate | 304 |

## B. TPC-H Workload Details

There is little work on using the standard TPC benchmarks in stream processing contexts. On the other hand, the linear road benchmark as the only standard and well-accepted benchmark in stream processing domain is not as mature as TPC benchmarks. Specifically, lack of standard query generator, detailed specification on stream load generation distribution model, limited number of complex queries, etc. make the accuracy investigation of the model using another workload a necessity.

We generated 35 executable queries based on the 13 TPC-H template queries. These queries registered against different arrival rates and their resource usage was logged. The number of collected traces and the window size for each query are shown in Table 5. For queries 7-9, 12-14, and 19, we limited the number of generated executable query to the value of 1, since the large window size allow the query to experience different selectivity ratio and resource usage spectrum.

**Table 5: 13 out of 22 TPC-H template queries constitute workload traces for model training and testing.**

| Query Temp. No. | Window Size (Seconds) | No. of generated executable query | No. of Traces |
| --- | --- | --- | --- |
| 1 | 120 | 5 | 594 |
| 2 | 810 | 6 | 2392 |
| 3 | 600 | 6 | 1284 |
| 5 | 2402 | 3 | 449 |
| 6 | 240 | 6 | 716 |
| 7 | 6005 | 1 | 312 |
| 8 | 6005 | 1 | 750 |
| 9 | 6005 | 1 | 562 |
| 10 | 600 | 2 | 409 |
| 12 | 1201 | 1 | 153 |
| 13 | 1500 | 1 | 326 |
| 14 | 1201 | 1 | 601 |
| 19 | 6005 | 1 | 234 |

Apart from window size, we needed to apply other minimal modifications to the TPC-H queries to make them executable in system X. These changes are shown in Table 6.

**Table 6: The changes applied to queries of TPC-H.**

| Index | Changes | Description | Applied to Queries No. |
| --- | --- | --- | --- |
| 1 | Using a built-in function for converting String to Date | The load generator doesn't support date data type, so that dates were sent as string and then converted to date in the middle of query execution. | 1,3,5,6,7,8, 10,12,14 |
| 2 | Using Date instead of Year as the system does not support function extract(). | A work-around is using user defined functions (UDFs), however the system is error prone with the usage of udfs. | 7,8,9 |
| 3 | Using LIKE clause for all string comparisons. | The equality and inequality operators of the language do not work for strings, so that we replace them with LIKE clause. | 2,3,5,7,8, 10,12,19 |
| 4 | Handling INTERVAL clause manually. | The INTERVAL clause of the language is limited and does not support the expected functionality of the queries. Thus, we applied the interval calculation manually before query execution. | 1,5,6,10,12 ,14 |

Among the modifications, we believe that the only change with the index number 2 does affect the resource usage of the query as it directly hits the selectivity ratio of the input tuples. However, the modified queries has a large window size (more than 1 hour) which let the input tuples gradually enter the windows and stay for a reasonable amount of time. Therefore, the queries are executed against various selectivity ratios which enable them to show all possible resource usage fluctuations. This also is not a major threat to the validity of the evaluation, since the focus of the study is the robust estimation of resource usage distribution.